# Coupling Machine Learning and Crop Modeling Improves Crop Yield Prediction in the US Corn Belt

Mohsen Shahhosseini[1], Guiping Hu[1*], Isaiah Huber[2], Sotirios V. Archontoulis[2],

[1] Department of Industrial and Manufacturing Systems Engineering, Iowa State University, Ames, Iowa, USA
[2] Department of Agronomy, Iowa State University, Ames, Iowa, USA
* Corresponding author e-mail: gphu@iastate.edu

## Abstract

This study investigates whether coupling crop modeling and machine learning (ML) improves corn yield predictions in the US Corn Belt. The main objectives are to explore whether a hybrid approach (crop modeling + ML) would result in better predictions, investigate which combinations of hybrid models provide the most accurate predictions, and determine the features from the crop modeling that are most effective to be integrated with ML for corn yield prediction. Five ML models (linear regression, LASSO, LightGBM, random forest, and XGBoost) and six ensemble models have been designed to address the research question. The results suggest that adding simulation crop model variables (APSIM) as input features to ML models can decrease yield prediction root mean squared error (RMSE) from 7 to 20%. Furthermore, we investigated partial inclusion of APSIM features in the ML prediction models and we found soil moisture related APSIM variables are most influential on the ML predictions followed by crop-related and phenology-related variables. Finally, based on feature importance measure, it has been observed that simulated APSIM average drought stress and average water table depth during the growing season are the most important APSIM inputs to ML. This result indicates that weather information alone is not sufficient and ML models need more hydrological inputs to make improved yield predictions.

## 1. Introduction

Advances in machine learning and simulation crop modeling have created new opportunities to improve prediction in agriculture[1–4]. These technologies have each provided unique capabilities and significant advancements in the prediction performance, however, they have been mainly assessed separately and there may be benefits integrating them to further increase prediction accuracy[5].

Simulation crop models predict yield, flowering time, and water stress using management, crop cultivar and environmental inputs and science-based equations of crop physiology, hydrology and soil C and N cycling[6–8]. In fact, these crop models are pre-trained using a diverse set of experimental data from various environments and are further refined (calibrated) for more accurate predictions in each study[9,10]. Numerous studies have used crop models for forecasting applications. For instance, Dumont et al.[11] compared the within-season yield predictive performance of two crop models, one model based on stochastically generated climatic data, and the other on mean climate data. The results show similar performance of both models with relative root mean square error (RRMSE) of 10% in 90% of the climatic situations. However, the model based on mean climate data had far less running time. Togliatti et al.[12] used APSIM maize and soybean to forecast phenology and yields with and without including weather forecast data. They found that inclusion of 7 to 14 day weather forecast did not improve end of season yield prediction accuracy. There are many other examples in the literature, in which crop modeling was used to forecast various aspects of the cropping system[13–15].

On the other hand, machine learning (ML) intends to make predictions by finding connections between input and response variables. Unlike simulation crop models, ML includes methods in which the system "learns" a transfer function to predict the desired output based on the provided inputs, rather than the researcher providing the transfer function. In addition, it is more easily applicable than simulation crop models as it does not require expert knowledge and user skills to calibrate the model, has lower runtimes, and less data storage constraints[8]. In recent years, there are several applications of ML algorithms to predict agronomic variables[16–21]. Drummond et al.[22] applied stepwise multiple linear regression (SMLR), projection pursuit regression (PPR), and several types of neural networks on a data set constructed with soil properties and topographic characteristics for 10 "site-years" with the purpose of predicting grain yields. They found that neural network models outperformed SMLR and PPR in every site-year. Khaki and Wang[23] designed residual neural network models to predict yield with prediction. Khaki et al.[24] developed a convolutional neural network – recursive neural network (CNN-RNN) framework to predict corn and soybean yields of 13 states in the US Corn Belt. Their model outperformed random forest, deep fully connected neural networks (DFNN), and least absolute shrinkage and selection operator (LASSO) models, achieving an RRMSE of 9% and 8% for corn and soybean prediction, respectively. Jiang et al.[25] devised a long short-term memory (LSTM) model that incorporates heterogeneous crop phenology, meteorology, and remote sensing data in predicting county-level corn yields. This model outperformed LASSO and random forest and explain 76% of yield variations across the Corn Belt. Mupangwa et al.[26] evaluated the performance of several ML models in predicting maize grain yields under conservation agriculture. The problem was formatted as a classification problem with the objective of labeling unseen observations' agro-ecologies (highlands or lowlands). They found that Linear discriminant analysis (LDA) performed better than other trained models, including logistic regression, K-nearest neighbor, decision tree, naïve Bayes, and support vector machines (SVM), with prediction accuracy of 61%.

We hypothesized that merging prediction tools, namely simulation crop models and machine learning models will improve prediction in agriculture. To our knowledge, there are no systematic studies in this area other than a few papers on combining crop models with simple regression. The main method has been the use of regression analysis to incorporate yield technology trends into the crop model simulations[27–30]. Some studies have used simulation crop model outputs as inputs to a multiple linear regression model and formed a hybrid simulation crop–regression framework to predict yields[31–34]. However, only two recent studies created hybrid simulation crop modeling–ML models for yield prediction. Everingham et al.[35] considered simulated biomass from the APSIM sugarcane crop model, seasonal climate prediction indices, observed rainfall, maximum and minimum temperature, and radiation as input variables of a random forest regression algorithm to predict annual variation in regional sugarcane yields in northeastern Australia. The results showed that the hybrid model was capable of making decent yield predictions explaining 67%, 72%, and 79% of the total variability in yield, when predictions are made on September 1st, January 1st, and March 1st, respectively. In another recent study, Feng et al.[36] claimed that incorporating machine learning with a biophysical model can improve the evaluation of climate extremes' impact on wheat yield in south-eastern Australia. To this end, they designed a framework that used the APSIM model outputs and growth stage-specific extreme climate events (ECEs) indicators to predict wheat yield using a random forest (RF) model. The developed hybrid APSIM + RF model outperformed the benchmark (hybrid APSIM + multiple linear regression (MLR)) and the APSIM model alone. The APSIM + RF introduced 19% and 33% improvements in the prediction accuracy of APSIM + MLR and APSIM alone, respectively. None of these studies compared the performance of various ML models and their ensembles in creating hybrid simulation crop modeling – ML frameworks and partial inclusion of the simulation crop modeling outputs is not studied in the literature.

The goal of this paper is to investigate the effect of coupling process-based modeling with machine learning algorithms towards improved crop yield prediction. The specific research objectives include:

1. Explore whether a hybrid approach (simulation crop modeling + ML) would result in better corn yield predictions in three major US Corn Belt states (Illinois, Indiana, and Iowa);
2. Investigate which combinations of hybrid models (various ML x crop model) provide the most accurate predictions;
3. Determine the features from the crop modeling that are most relevant for use by ML for corn yield prediction.

Figure 1 depicts the conceptual framework of this paper.

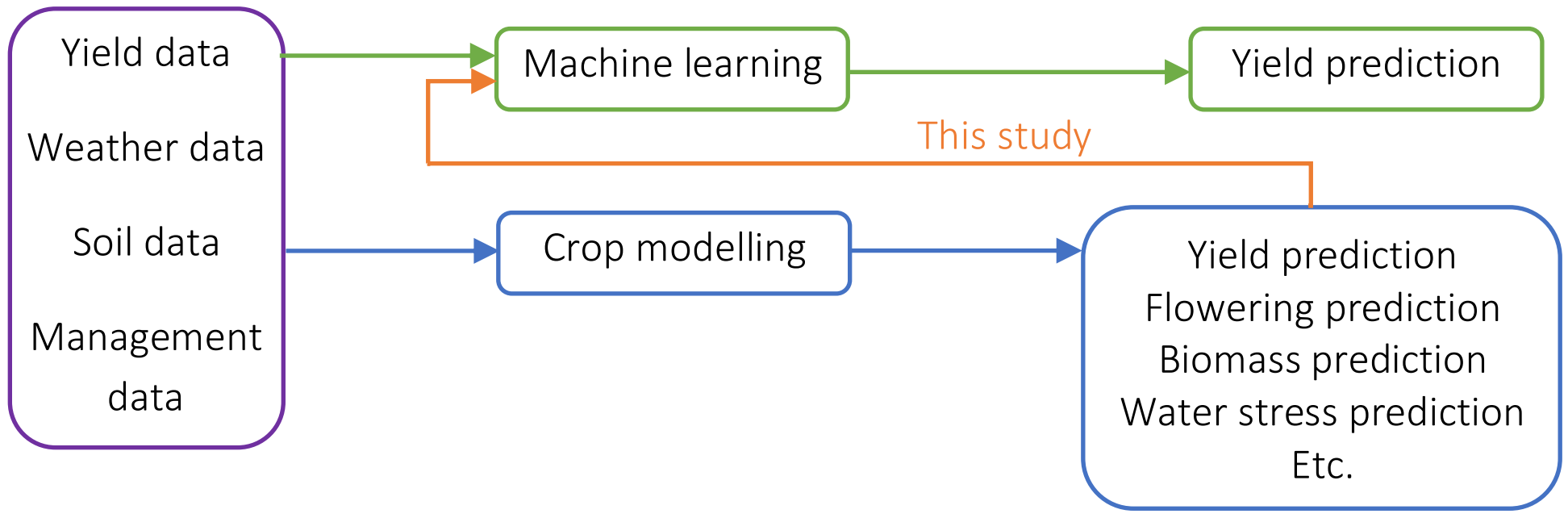

*Figure 1: Conceptual framework of this study's objective. Note that yield data are not an input in crop modeling. However, yield data are indirectly used to test and improve crop model predictions as needed.*

The remainder of this paper is organized as follows. Section 2 describes the methodology and the materials used in this study, and Section 3 presents and discusses the results and the possible improvements. Section 4 discusses the analysis and findings and finally, Section 5 concludes the paper.

## 2. Materials and Methods

Since the main objective is to evaluate the performance of a hybrid simulation-machine learning framework in predicting corn yield, this section is split into two parts. The first describes the Agricultural Production Systems sIMulator (APSIM) and the second the Machine learning (ML) algorithms. Each of them explains the details of the prediction/forecasting framework, including the inputs to the models, the data processing tasks, the details of selected predictive models, and evaluation metrics used to compare the results, for simulation and machine learning.

## 2.1. Agricultural Production Systems sIMulator (APSIM)

### 2.1.1. APSIM run details

The Agricultural Production Systems sIMulator[37] (APSIM) is an open source advanced simulator of cropping systems. It includes many crop models along with soil water, C, N, crop residue modules, which all interact on a daily time step. In this project, we used the APSIM maize version 7.9 and in particular the calibrated model version for US Corn Belt environments as outlined by Archontoulis et al.[1] that includes simulation of shallow water tables and inhibition of root growth due to excess water stress[38] and waterlogging functions[39]. Within APSIM we used the following modules: maize[40], SWIM soil water[41], soil N and carbon[42], surface residue[42,43], soil temperature[44] and various management rules to account for tillage and other management operations. The crop models simulate potential biomass production based on a combined radiation and water use efficiency concept. This potential is reduced to attainable yields by incorporating water and nitrogen limitation to crop growth (For additional information, we refer to www.apsim.info).

To run APSIM across the three states Illinois, Indiana, and Iowa, we used the parallel system for integrating impact models and sectors (pSIMS) software[45]. pSIMS is a platform for generating simulations and running point-based agricultural models across large geographical regions. The simulations used in this study were created on a 5-arcminute grid across Iowa, Illinois and Indiana considering only cropland area when creating soil profiles. Soil profiles for these simulations were created from Soil Survey Geographic database (SSURGO)[46], a soil database based off of soil survey information collected by the National Cooperative Soil Survey. Climate information used by the simulations came from a synthetic weather data set called "IEM Reanalysis", which was engineered at Iowa Environmental Mesonet (mesonet.agron.iastate.edu). This database is developed from a combination of several weather sources. The temperature data comes from National Weather Service Cooperative Observer Program (NWS COOP) observers (www.weather.gov/coop). The precipitation data is derived from radar-based estimates of National Oceanic and Atmospheric Administration Multi-Radar / Multi-Sensor System (NOAA MRMS) (www.nssl.noaa.gov/projects/mrms), Oregon State's PRISM data set (https://prism.oregonstate.edu/), and NWS COOP reports. Finally, the radiation data comes from NASA POWER (power.larc.nasa.gov). The synthetic product was tested against point weather stations and proved accurate (see more information here: https://crops.extension.iastate.edu/facts/weather-tool). Current management APSIM model input databases include changes in plant density, planting dates, cultivar characteristics and N fertilization rate to corn from 1984 to 2019. Planting date and plant density data derived from USDA-NASS[47]. Cultivar traits data derived through regional scale model calibration. N fertilizer data derived from a combined analysis of USDA-NASS[47] and Cao et al.[48] including N rates to corn by county and by year. Over the historical period, 1984-2019, APSIM captured 78% of the variability in the NASS yields having a RMSE of 1 Mg/ha and RRMSE of 10% (See Figure 2). This version of the model is used to provide outputs to the machine learning.

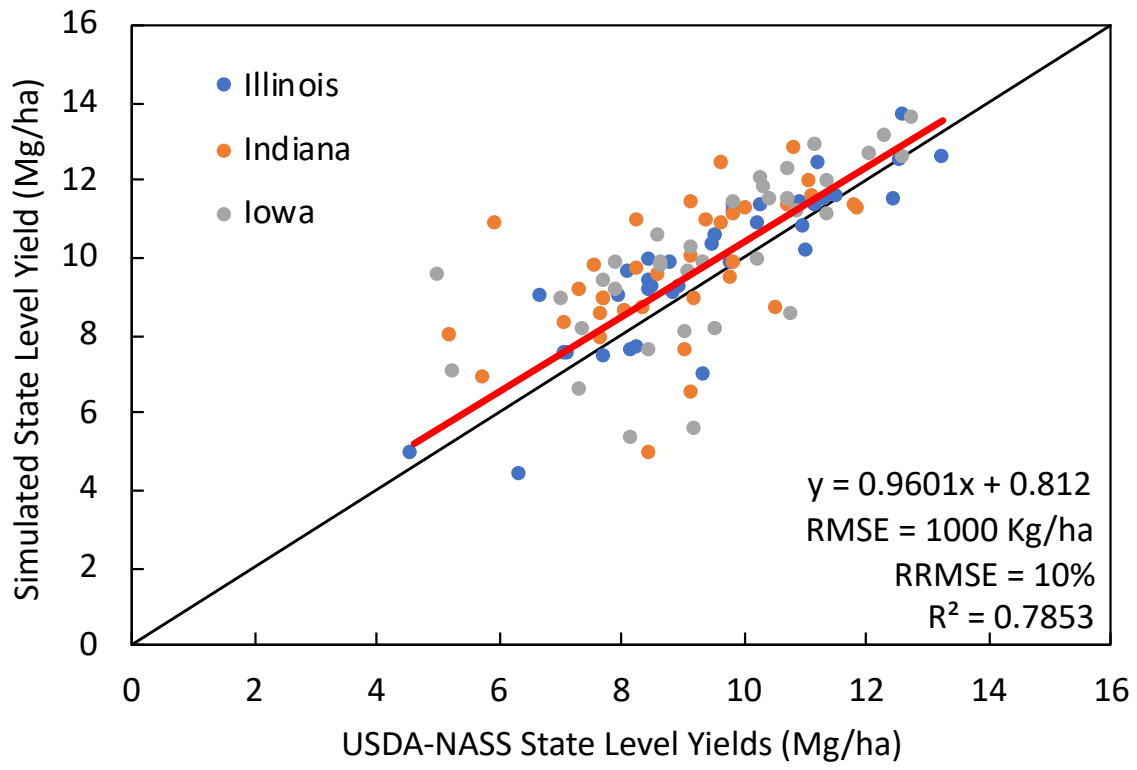

*Figure 2: Measured (USDA-NASS) corn yields vs. simulated corn yields at the state level from 1984 to 2019 using the pSIMS-APSIM framework.*

### 2.1.2. APSIM output variables used as inputs to ML models

The first step to combine the developed data set with APSIM variables was to extract all APSIM simulations from its outputs and prepare the obtained data to be added to the mentioned data set. The APSIM outputs include 22 variables (the details are presented in Table 1). The granularity level for the APSIM variables was different from USDA obtained data, as the APSIM variables made at 5 arc (approximately 40 fields within a county). Therefore, to calculate a county-level value for each of them, the median of all corresponding values is used. The reason to use median instead of a simple average is to reduce the impact of outliers on yields. Among the 40 fields/county * 300 counties * 35 yields there were some model failures or zero yields that bias the county level yield predictions.

*Table 1: Description of all APSIM outputs added to the developed data set for building ML models*

| | Acronym | Description |
|---|---|---|
| 1 | Crop Yield | Crop yield (kg/ha) |
| 2 | Biomass | Crop above ground biomass (kg/ha) |
| 3 | Root Depth | Maximum root depth (mm) |
| 4 | Flower Date | Flowering time (doy) |
| 5 | Maturity Date | Maturity time (doy) |
| 6 | LAI maximum | Maximum leaf area index (m2/m2) |
| 7 | ET Annual | Actual evapotranspiration (mm) |
| 8 | Crop Transpiration | Crop transpiration (mm) |
| 9 | Total Nupt | Above ground crop N uptake (Kg N/ha) |
| 10 | Grainl Nupt | Grain N uptake (kg N/ha) |
| 11 | Avg Drought Stress | Average drought stress on leaf development (0-1) |
| 12 | Avg Excessive Stress | Average excess moisture stress on photosynthesis (0-1) |
| 13 | Avg N Stress | Average N stress on grain growth (0-1) |
| 14 | Avg WT Inseason | Depth to water table during the growing season (mm) |
| 15 | Runoff Annual | Runoff (mm) |
| 16 | Drainage | Drainage from tiles and below 1.5 m (mm) |
| 17 | Gross Miner | Soil gross N mineralization (kg N/ha) |
| 18 | Nloss Total | Total N loss (denitrification and leaching) kg N/ha |
| 19 | Avg WT | Depth to water table during the entire year (mm) |
| 20 | SWtoDUL30Inseason | Growing season average soil water to field capacity ratio at 30 cm |
| 21 | SWtoDUL60Inseason | Same as above but at 60 cm |
| 22 | SWtoDUL90Inseason | Same as above but at 90 cm |

All 22 APSIM output values were prepared and added to the developed data set. The pre-processing tasks done for APSIM data were:

- Imputing zero values with the average of other values of the same feature
- Removing rows with missing values
- Normalizing the data to be between 0 and 1
- Cross-referencing the new data with the developed data set

Then, all feature selection procedures explained in section 2.2.2 were executed on the newly created data set to keep only the variables that carry the most relevant information for the prediction task.

The developed data set considers data from 1984 to 2018. The data from three years, namely 2012, 2017, and 2018 are in turn considered as the test data and for each scenario, the training data is set to be the data from the other years. In essence, we considered average to wet years (2017 and 2018) and an extremely dry year (2012) as the test years to assess the model performance in all situations.

## 2.2. Machine Learning (ML)

The machine learning models are developed using a data set spanning from 1984 to 2018 to predict corn yield in three US Corn Belt states (Illinois, Indiana, and Iowa). The data set is comprised of the environment (soil and weather) and management as input variables, and actual corn yields for the period under study as the target variable. The input data are comprised of weather, management, and soil data[1]. Environment data includes several soil parameters at a 5 km resolution[46] and weather data.

### 2.2.1. Data set

The county-level historical corn yields were downloaded from the USDA National Agricultural Statistics Service[47] for years 1984-2018. A data set including observed information of the environment, management, and yields was developed, which consists of 10,016 observations of yearly average corn yields for 293 counties. The factors that mainly affect crop yields are alleged to be the environment, genotype, and management. To this end, weather and soil as environmental features and plant population and planting progress as management features were included in the data set. It should be noted that data preprocessing has been designed to address the increasing trends in yields due to technological and genotypic advances over the years[49,50]. This is mainly due to that there is no publicly available genotype data set. The data set with 598 variables (including target variable) are described below.

- *Plant population:* one feature describing the plant population per year and per state measured in plants per square meter, obtained from USDA-NASS[47]
- *Planting progress (planting date):* 52 features describing the weekly cumulative percentage of corn planted within each state[47]
- *Weather*: Five weather variables accumulated weekly (260 features), obtained from Iowa Environmental Mesonet
    1. Daily minimum air temperature in degrees Celsius
    2. Daily maximum air temperature in degrees Celsius

3. Daily total precipitation in millimeters per day
4. Growing degree days in degrees Celsius (base 10 ceiling 30)
5. Daylight average incident shortwave radiation in Megajoules per square meter

- *Soil*: The soil features soil organic matter, sand content, clay content, soil pH, soil bulk density, wilting point, field capacity, and saturation point, were considered in this study. Different values for different soil layers were used as the features mentioned above change across the soil profile. Consequently, 180 features for soil characteristics of the locations under study were obtained from the Web Soil Survey[46]
- *Corn Yield*: Yearly corn yield data in bushel per acre, collected from USDA-NASS[47]

### 2.2.2. Data pre-processing

Several pre-processing tasks were conducted to ensure the data is prepared for fitting machine learning models. Since it is favorable for some machine learning models especially weighted ensemble models for the data input to have similar ranges, the first pre-processing task was to scale the input data between 0 and 1 using min-max scaling. The most common scaling methods include min-max scaling and normalization, from which min-max scaling is selected as it keeps the distributions of the input variables. The next pre-processing tasks include adding yearly trends, cumulative weather feature construction, and feature selection.

- *Add yearly trends feature*

Figure 3 suggests an increasing trend in the yields over time. It is evident that there is no input feature in the developed data set that can explain this observed increasing trend in the corn yields. This trend is commonly described as the effect of technological gains over time, such as improvements in genetics (cultivars), management[51], equipment, and other technological advances[52,53].

Therefore, to account for the trend as mentioned above, the following actions were taken.

A new feature (yield_trend) was constructed that only explained the observed trend in corn yields. For building this new feature, a linear regression model was built for each location as the trends for each site tend to be different. The year ($YEAR$) and yield ($Y$) features formed the independent and dependent variables of this linear regression model, respectively. Then the predicted value for each data point ($\widehat{Y}$) is added as a new input variable that explains the increasing annual trend in the target variable. Only training data was used for fitting this linear regression model and the corresponding values of the newly added feature for the test set is set to be the predictions made by this model for the data of that year ($\widehat{Y}_{i,test} = b_{0_i} + b_{1_i} YEAR_{i,test}$). The following equation shows the trend value ($\widehat{Y}_i$) calculated for each location ($i$), that is added to the data set as a new feature.

$$\widehat{Y}_i = b_{0_i} + b_{1_i} YEAR_i \quad (1)$$

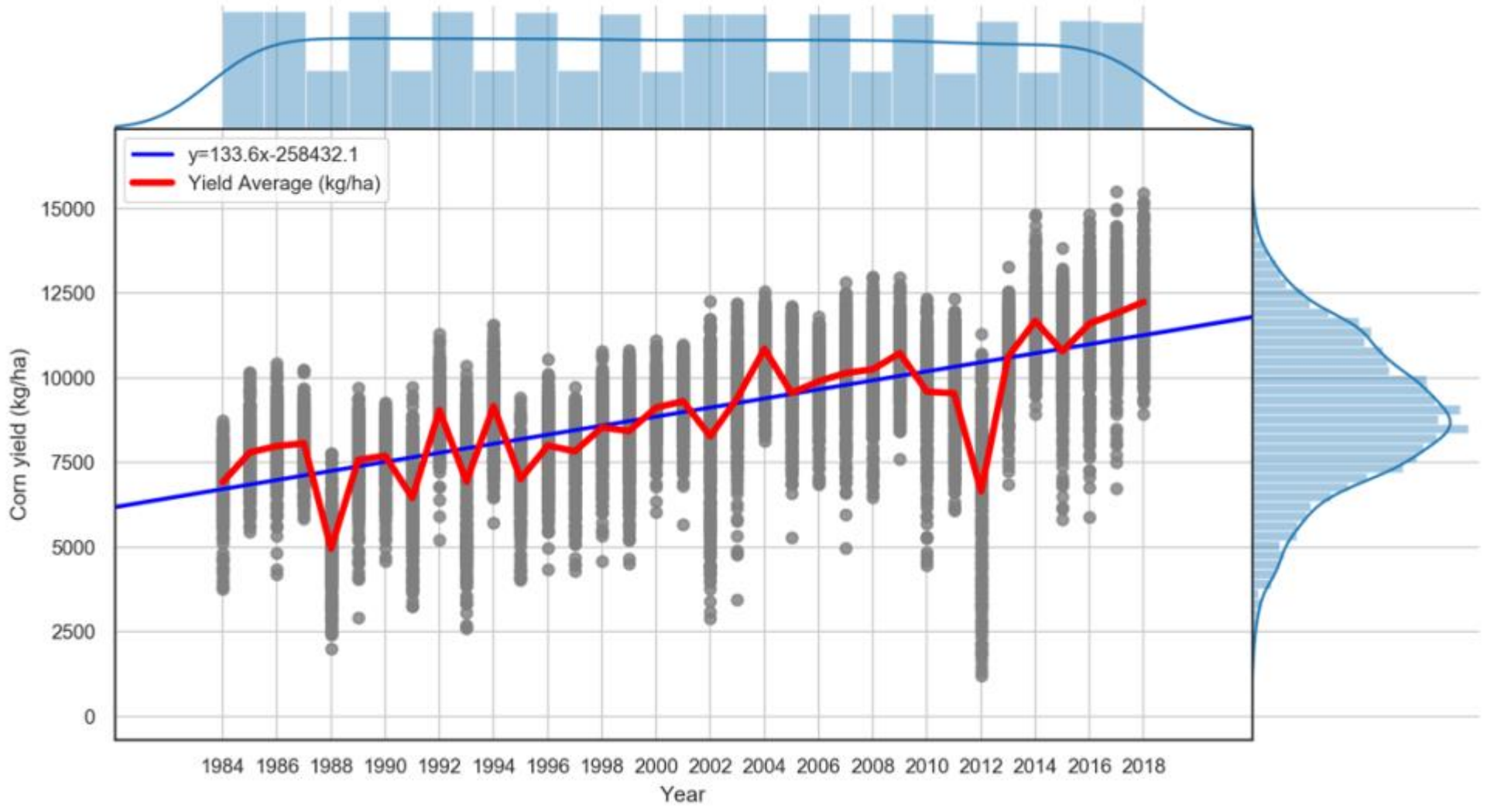

*Figure 3: Plotting aggregated annual yields for all locations under study and the average yields per year The figure shows the increase in yield with time and the distribution of residuals around the regression*

- *Aggregated and cumulative weather feature construction*

To provide more climate information for the machine learning models, additional weather features were constructed that include cumulated values of the existing weather features. The aggregated precipitation, growing degree days, and shortwave radiation features are computed from summation of weather features, while the aggregated minimum and maximum temperature features come from average of the existing values. There are two sets of new cumulative weather features: Quarterly weather features (20 features), and cumulative quarterly weather features (15 features)

- *Feature selection*

Since the data developed data set has a large number of input variables and is prone to overfitting, feature selection becomes necessary to build generalizable machine learning models. A two-stage feature selection procedure was performed to select the most essential features in the data set and prevent the machine learning models from overfitting on the highly dimensional training data. The two steps to perform feature selection were feature selection based on expert knowledge, and permutation feature selection using random forest.

*i. Feature selection based on expert knowledge*

Using expert knowledge, weather features were reduced by removing features for the period between the end of harvesting and the beginning of next year's planting. Additionally, the number of planting progress features were lowered by eliminating the cumulative planting progress for the weeks before planting, as they did not include useful information. The feature selection based on expert knowledge could reduce the number of features from 550 to 387.

*ii. Permutation feature selection with random forest*

Strobl[54] pointed out that the default random forest variable importance (impurity-based) is not reliable when dealing with situations where independent variables have different scales of measurement or different number of categories. This is specifically important for biological and genomic studies where independent variables are often a combination of categorical and numeric features with varying scales. Therefore, to overcome this bias and find decisive importance of input features, permutation feature importance is decided to be used[55].

Permutation feature importance measures the importance of an input feature by calculating the decrease in the model's prediction error when one feature is not available[56]. To make the unavailability of one feature possible, each feature is permuted in the validation or test set, that is, its values are shuffled, and the effect of this permutation on the quality of the predictions is measured. Specifically, if permutation increases the model error, the permuted feature is considered important, as the model relies on that feature for prediction. On the other hand, if permutation does not change the prediction error significantly, the feature is thought to be unimportant, as the model ignores it for making the prediction[57].

The second stage of feature selection and likely the most effective one, includes fitting a random forest model with 100 number of trees as the base model and calculating permutation importance of input features with 10 times of repetition and considering a random 10-fold cross-validation schema. It should be noted that the number of trees hyperparameter of this random forest model is tuned using a 10-fold cross-validation. Afterward, the top 80 input features were selected in the second stage of feature selection.

### 2.2.3. Model selection

Tuning hyperparameters of machine learning models and selecting best models with optimal hyperparameter values is necessary to achieve high prediction accuracies. Cross-validation is commonly used to evaluate the predictive performance of fitted models by dividing the training set to train and validation subsets. Here, we use a random 10-fold cross-validation method to tune the hyperparameter of ML models.

Grid search is an exhaustive search method that tries all the possible combinations of hyperparameter settings to find the optimal selection. It is both computationally expensive and generally dependent on the initial values specified by the user. However, Bayesian search addresses both issues and is capable of tuning hyperparameters faster and using a continuous range of values.

Bayesian search assumes an unknown underlying distribution and tries to approximate the unknown function with surrogate models such as Gaussian process. Bayesian optimization incorporates prior belief about the underlying function and updates it with new observations. This makes tuning hyperparameters faster and ensures finding an acceptable solution, given that enough number of observations are observed. In each iteration, Bayesian optimization gathers observations with the highest amount of information and intends to make a balance between exploration (exploring uncertain hyperparameters) and exploitation (gathering observations from hyperparameters close to the optimum)[58]. That being so, to tune hyperparameters, Bayesian search with 20 iterations was selected as the search method under 10-fold cross-validation procedure.

## 2.3. Predictive models

In this study, we combine diverse models in different ways and create ensemble models to make a robust and precise machine learning model. One prerequisite for creating well-performing ensemble models is to show a particular element of diversity in the predictions of base learners as well as preserve excellent performance individually[59]. Thus, several base learners made with different procedures were selected and trained, including linear regression, LASSO regression, Extreme Gradient Boosting (XGBoost), Light Gradient Boosting Machine (LightGBM), and random forest. Moreover, an average weighted ensemble that assigns equal weights to all base learners is the simplest ensemble model created. Additionally, optimized weighted ensemble method proposed in Shahhosseini et al.[60] was applied here to test its predictive performance. Several two-level stacking ensembles, namely stacked regression, stacked LASSO, stacked random forest, and stacked LightGBM, were built, which are expected to demonstrate excellent performance. The details of each model can be found at Shahhosseini et al.[61].

### 2.3.1. Linear regression

Linear regression intends to predict a measurable response using multiple predictors. It assumes the existence of a linear relationship between the predictors and response variable, normality, no multicollinearity, and homoscedasticity[62].

### 2.3.2. LASSO regression

LASSO is a regularization method that is equipped with in-built feature selection. It can exclude some variables by setting their coefficient to zero[62]. Specifically, it adds a penalty term to the linear regression loss function, which can shrink coefficients towards zero (L1 regularization)[63].

### 2.3.3. XGBoost and LightGBM

XGBoost and LightGBM are two implementations of gradient boosting tree-based ensemble methods. These types of ensemble methods make predictions sequentially and try to combine weak predictive tree models and learn from their mistakes. XGBoost was proposed in 2016 with new features, such as handling sparse data, and using an approximation algorithm for a better speed[64], while LightGBM was published in 2017 by Microsoft, with improvements in performance and computational time[65].

### 2.3.4. Random forest

Random forest is built on the concept of bagging, which is another tree-based ensemble model. Bagging tries to reduce prediction variance by averaging predictions made by sampling with replacement[66]. Random forest adds a new feature to bagging, which is randomly choosing a random number of features and constructing a tree with them and repeating this procedure many times and eventually averaging all the predictions made by all trees[59]. Therefore, random forest addresses both bias and variance components of the error and is proved to be powerful[67].

### 2.3.5. Optimized weighted ensemble

An optimization model was proposed in Shahhosseini et al.[60], which accounts for the tradeoff between bias and variance of the predictions, as it uses mean squared error (MSE) to form the objective function for the optimization problem[68]. In addition, out-of-bag predictions generated by $k$-fold cross-validation are used as emulators of unseen test observations to create the input matrices of the optimization problem, which are out-of-bag predictions made by each base learner. The optimization problem, which is a nonlinear convex problem, is as follows.

$$
\begin{aligned}
&Min\ \ \frac{1}{n}\sum_{i=1}^{n}\left(y_i - \sum_{j=1}^{k} w_j \hat{y}_{ij}\right)^2 \\
&s.t. \\
&\quad \sum_{j=1}^{k} w_j = 1, \\
&\quad w_j \geq 0, \quad \forall j = 1, \dots, k.
\end{aligned} \tag{3}
$$

where $w_j$ is the weights corresponding to base model $j$ ($j = 1, \dots, k$), $n$ is the total number of instances, $y_i$ is the actual value of observation $i$, and $\hat{y}_{ij}$ is the prediction of observation $i$ by base model $j$.

### 2.3.6. Average weighted ensemble

Average weighted ensemble, which we call "average ensemble", is a simple average of out-of-bag predictions made by each base learner. The average ensemble can perform well when the base learners are diverse enough[59].

### 2.3.7. Stacked generalization

Stacked generalization tries to combine multiple base learners by performing at least one more level of learning task, that uses out-of-bag predictions for each base learner as inputs, and the actual target values of training data as outputs[69]. The out-of-bag predictions are generated through a $k$-fold cross-validation and have the same size of the original training set[70]. The steps to design a stacked generalization ensemble are as follows.

a) Learn first-level machine learning models and generate out-of-bag predictions for each of them by using $k$-fold cross-validation.
b) Create a new data set with out-of-bag predictions as the input variables and actual response values of data points in the training set as the response variable.
c) Learn a second-level machine learning model on the created data set and make predictions for unseen test observations.

Considering four predictive models as the second-level learners, four stacking ensemble models were created, namely stacked regression, stacked LASSO, stacked random forest, and stacked LightGBM.

## 2.4. Performance metrics

To evaluate the performance of the developed machine learning models, three statistical performance metrics were used.

- Root Mean Squared Error (RMSE): the square root of the average squared deviation of predictions from actual values[71].
- Relative Root Mean Squared Error (RRMSE): RMSE normalized by the mean of the actual values
- Mean Bias Error (MBE): a measure that describes the average bias in the predictions.
- Coefficient of determination ($R^2$): the proportion of the variance in the dependent variable that is explained by independent variables.

These metrics together provide estimates of the error (RMSE, RRMSE, MBE) and of the variance explained by the models ($R^2$).

# 3. Results

## 3.1. Numerical results of hybrid simulation – ML framework

Table 2 shows the test set prediction errors of the 11 developed ML models for the benchmark (the case that no APSIM variable is added to the data set) and the hybrid simulation-ML (where all 22 APSIM outputs are added to the data set) cases. The relative RMSE (RRMSE) is calculated using the average corn yield value of the test set (see Table S1 in the supplementary material). Adding APSIM variables as input features to ML models improved the performance of the 11 developed ML models. In terms of RMSE, the hybrid model boosted ML performance up to 27%. In addition, comparing the lowest prediction errors (RMSE) of the benchmark and the hybrid scenario, we found that the use of hybrid models achieved 8%-9% better corn yield predictions.

*Table 2: Test set prediction errors for years 2017 and 2018 of ML models for benchmark and hybrid cases*

| *ML model* | *Benchmark (no APSIM variable)* | | | | *Hybrid simulation – ML (all 22 APSIM variables included)* | | | | *% decrease in RMSE* |
|---|---|---|---|---|---|---|---|---|---|
| | *RMSE (kg/ha)* | *RRMSE (%)* | *MBE (kg/ha)* | *$R^2$ (%)* | *RMSE (kg/ha)* | *RRMSE (%)* | *MBE (kg/ha)* | *$R^2$ (%)* | *%* |
| Test set: 2018 | | | | | | | | | |
| LASSO | 1160 | 9.5% | 559 | 24.5% | 1094 | 8.9% | 206 | 32.8% | 5.7% |
| XGBoost | 1482 | 12.1% | -879 | -23.3% | 1172 | 9.6% | -581 | 22.8% | 20.9% |
| LightGBM | 1067 | 8.7% | -549 | 36.1% | 883 | 7.2% | -89 | 56.2% | 17.3% |
| Random forest | 1259 | 10.3% | -717 | 11.1% | 1055 | 8.6% | -567 | 37.5% | 16.1% |
| Linear regression | 1095 | 8.9% | 589 | 32.7% | 955 | 7.8% | 100 | 48.8% | 12.7% |
| Optimized weighted ens. | 1033 | 8.4% | -485 | 40.1% | 909 | 7.4% | -192 | 53.6% | 12.0% |
| Average ensemble | 959 | 7.8% | -200 | 48.4% | 938 | 7.7% | -186 | 50.7% | 2.2% |
| Stacked regression ens. | 1140 | 9.3% | -705 | 27.1% | 943 | 7.7% | -23 | 50.1% | 17.3% |
| Stacked LASSO ensemble | 1128 | 9.2% | -685 | 28.5% | 941 | 7.7% | -29 | 50.3% | 16.6% |
| Stacked Random f. ens. | 1363 | 11.1% | -355 | -4.2% | 1002 | 8.2% | 49 | 43.6% | 26.5% |
| Stacked LightGBM ens. | 1365 | 11.2% | -366 | -4.6% | 995 | 8.1% | 43 | 44.4% | 27.1% |
| Test set: 2017 | | | | | | | | | |
| LASSO | 835 | 7.0% | -192 | 67.0% | 771 | 6.5% | 63 | 71.9% | 7.6% |
| XGBoost | 957 | 8.0% | -256 | 56.7% | 946 | 7.9% | -236 | 57.7% | 1.2% |
| LightGBM | 914 | 7.7% | -437 | 60.5% | 916 | 7.7% | -64 | 60.3% | -0.2% |
| Random forest | 1004 | 8.4% | -544 | 52.3% | 841 | 7.1% | -276 | 66.5% | 16.2% |
| Linear regression | 858 | 7.2% | -333 | 65.2% | 830 | 7.0% | 271 | 67.4% | 3.3% |
| Optimized weighted ens. | 885 | 7.4% | -404 | 62.9% | 787 | 6.6% | -8 | 70.7% | 11.1% |
| Average ensemble | 859 | 7.2% | -352 | 65.1% | 762 | 6.4% | -48 | 72.5% | 11.3% |
| Stacked regression ens. | 940 | 7.9% | -495 | 58.2% | 810 | 6.8% | 96 | 68.9% | 13.8% |
| Stacked LASSO ensemble | 935 | 7.9% | -486 | 58.7% | 809 | 6.8% | 100 | 69.0% | 13.4% |
| Stacked Random f. ens. | 993 | 8.3% | -331 | 53.4% | 888 | 7.5% | 63 | 62.7% | 10.6% |
| Stacked LightGBM ens. | 921 | 7.7% | -288 | 59.8% | 838 | 7.0% | 98 | 66.8% | 9.1% |

Looking at the average test results (Figure 4), it can be observed that adding APSIM features makes improvements to all designed ML models. Moreover, considering the smallest decrease in the prediction error (RRMSE) which is the worst-case scenario and is obtained by LASSO model, the hybrid model still is proved to be better than the benchmark. Another observation is the superiority of weighted ensemble models compared to other ML models. It should be noted that the negative $R^2$ value of some models (XGBoost, Stacked Random forest, and Stacked LightGBM) when having no APSIM variables shows that this models' predictions are worse than taking the mean value as the predictions.

On average, stacked ensemble models benefit the most from inclusion of APSIM outputs in predicting corn yields. Besides, considering Mean Bias Estimate (MBE) values of the ML models, we can observe that all ML models presented less biased predictions after having APSIM information in their inputs and it seems that inclusion of APSIM variables helped reducing the prediction bias significantly.

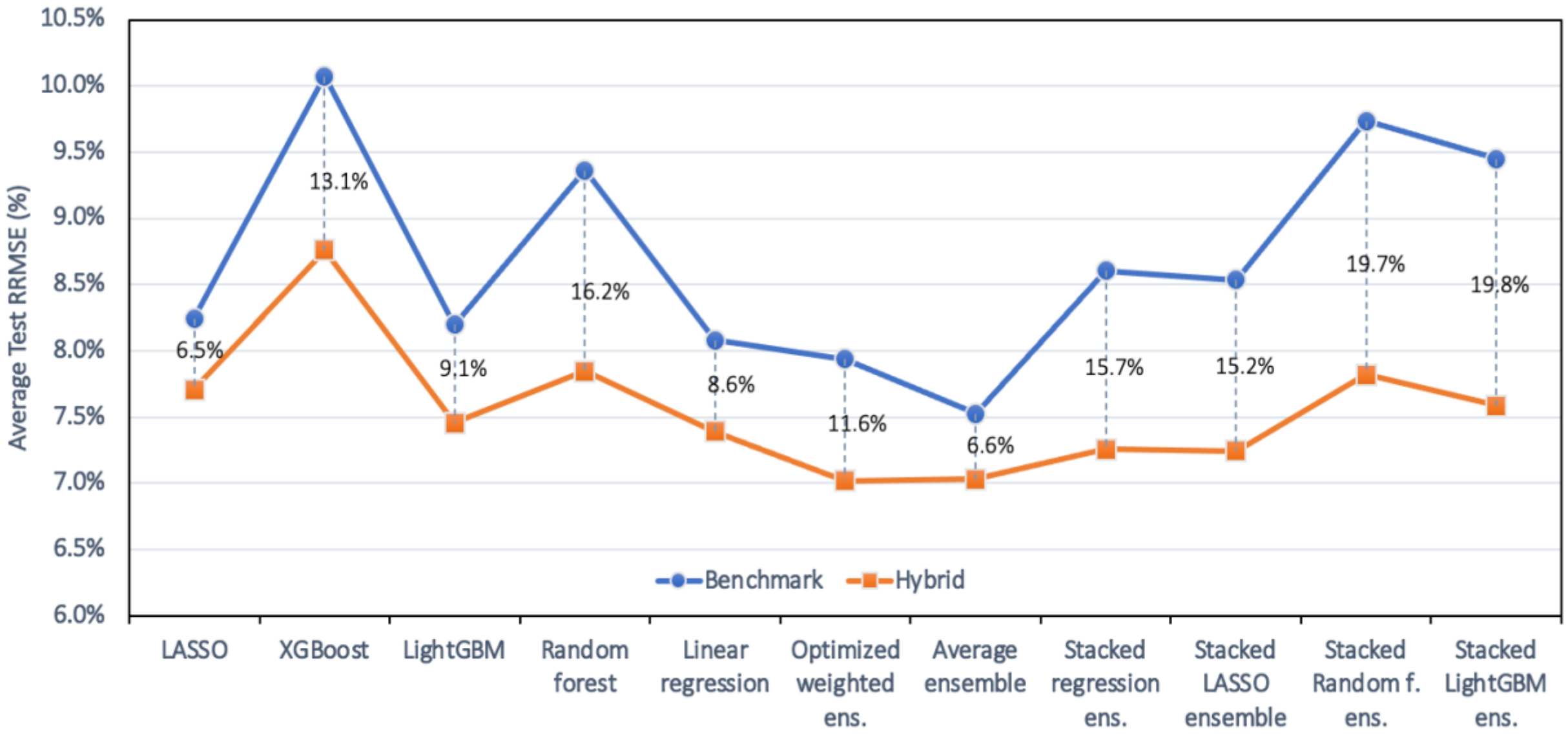

*Figure 4: Comparing average test RRMSE of benchmark and hybrid developed ML models. Data is averaged over the years 2017 and 2018*

Figure 5 illustrates the goodness of fit of some of the designed ML models for two benchmark and hybrid cases for the test year 2018. As mentioned above, the advantage of including APSIM variables in the machine learning algorithms is the better distribution of the residuals (deviation from the 1:1 line) which decreased overall prediction bias. See Figure S1 from supplementary material for additional summary statistics.

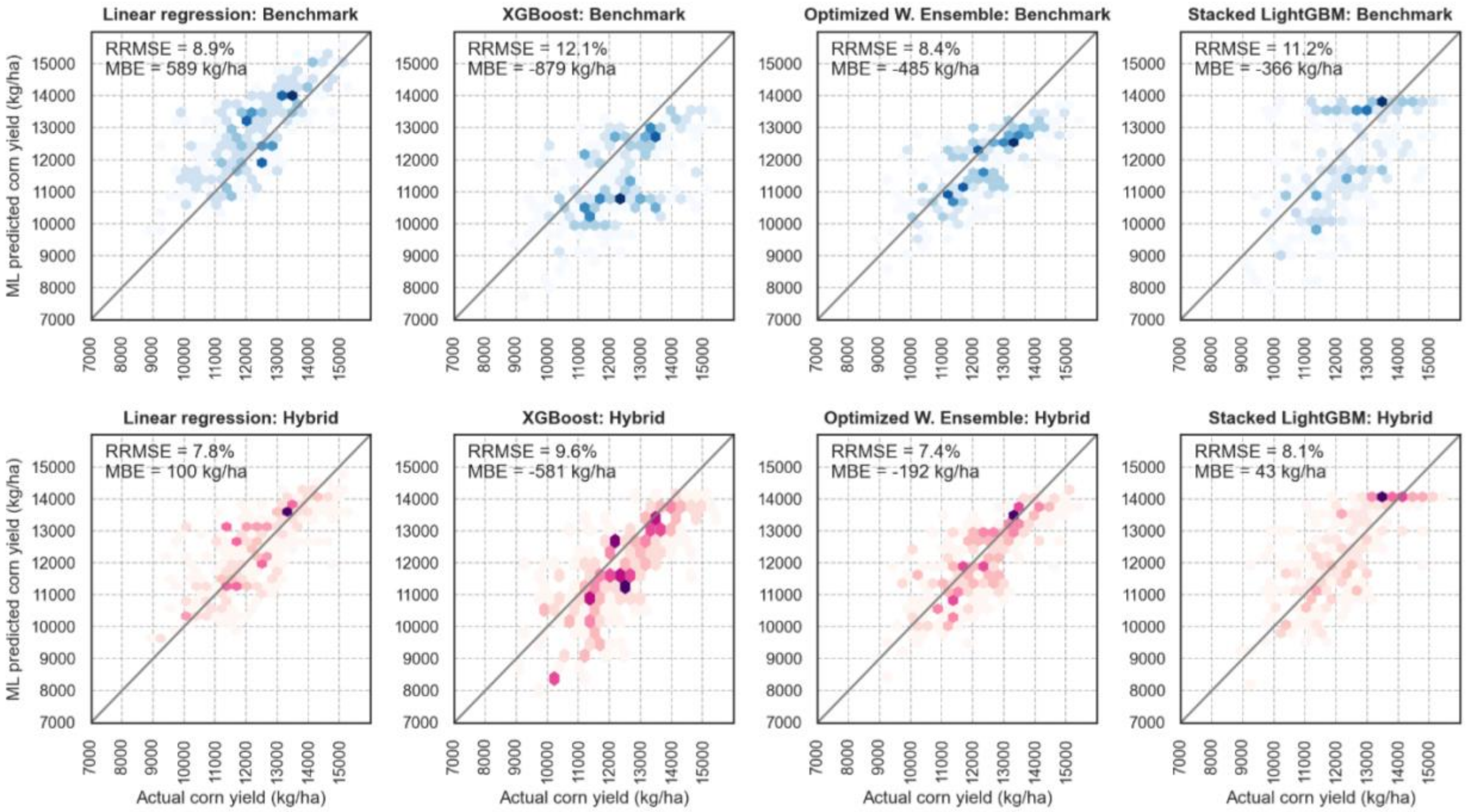

*Figure 5: X-Y plots of selected designed ML models for benchmark (top) and hybrid model (bottom) cases for test year 2018. The intensity of the colors shows the accumulation of the data points*

## 3.2. Models performance on an extreme weather year (2012)

To assess the performance of the trained models on an extreme weather year, here the data from the year 2012, which was an exceptionally dry year, is considered as unseen test observations and the quality of the predictions made by the benchmark and the hybrid models are compared.

Table 3 demonstrates lower prediction accuracy of the models in year 2012 (extreme dry year) compared to average to wet years model predictions (2017 and 2018, see Table 2). This result was consistent for both ML and hybrid models. However, the hybrid model managed to provide improvements over the benchmark in the 2012 year. This was ranging from 5% to 43% decrease in the prediction RMSE. Comparing the best model of the benchmark (LightGBM) with the best model of the hybrid scenario (Stacked regression ensemble), we observed that the use of hybrid model provided 22% better predictions.

*Table 3: Test set prediction errors of ML models for benchmark and hybrid cases when considering an extreme weather year (2012) – The average yield of the year 2012 is 6646 kg/ha*

| *ML model* | *Benchmark (no APSIM variable)* | | | | *Hybrid simulation – ML (all 22 APSIM variables included)* | | | | *% decrease in RMSE* |
|---|---|---|---|---|---|---|---|---|---|
| | *RMSE (kg/ha)* | *RRMSE (%)* | *MBE (kg/ha)* | *$R^2$ (%)* | *RMSE (kg/ha)* | *RRMSE (%)* | *MBE (kg/ha)* | *$R^2$ (%)* | *%* |
| Test set: 2012 | | | | | | | | | |
| LASSO | 4311 | 64.9% | 3775 | -221.3% | 3160 | 47.5% | 2519 | -72.6% | 26.7% |
| XGBoost | 3664 | 55.1% | 3102 | -132.1% | 2602 | 39.2% | 1942 | -17.1% | 29.0% |
| LightGBM | 3245 | 48.8% | 2671 | -82.0% | 2608 | 39.2% | 2016 | -17.5% | 19.6% |
| Random forest | 3782 | 56.9% | 3368 | -147.3% | 3591 | 54.0% | 3157 | -122.9% | 5.0% |
| Linear regression | 4869 | 73.3% | 4450 | -309.8% | 2784 | 41.9% | 2144 | -34.0% | 42.8% |
| Optimized weighted ens. | 3380 | 50.9% | 2818 | -97.5% | 2664 | 40.1% | 2066 | -22.7% | 21.2% |
| Average ensemble | 3946 | 59.4% | 3473 | -169.2% | 2908 | 43.8% | 2356 | -46.2% | 26.3% |
| Stacked regression ens. | 3398 | 51.1% | 2816 | -99.6% | 2545 | 38.3% | 1894 | -12.0% | 25.1% |
| Stacked LASSO ensemble | 3403 | 51.2% | 2835 | -100.1% | 2561 | 38.5% | 1925 | -13.4% | 24.7% |
| Stacked Random f. ens. | 3289 | 49.5% | 2668 | -87.0% | 2571 | 38.7% | 1968 | -14.3% | 21.8% |
| Stacked LightGBM ens. | 3462 | 52.1% | 2934 | -107.2% | 2588 | 38.9% | 1995 | -15.8% | 25.2% |

## 3.3. Partial inclusion of APSIM variables

This section investigates the effect of partial inclusion of APSIM variables considering three different scenarios for the test year 2018 (see Table 4). The scenarios are (1) include only phenology-related APSIM variables (silking date and physiological maturity date); (2) include only crop-related APSIM variables (crop yield, biomass, maximum rooting depth, maximum leaf area index, cumulative transpiration, crop N uptake, grain N uptake, season average water stress (both drought and excessive water), and season average nitrogen stress), and (3) include soil and weather-related APSIM variables (annual evapotranspiration, growing season average depth to the water table, annual runoff, annual drainage, annual gross N mineralization, total N loss that accounts for leaching and denitrification, annual average water table depth, ratio of soil water to field capacity during the growing season at 30, 60, and 90 cm profile depth). When including only phenology-related APSIM variables, results demonstrate that stacked regression ensemble model makes the best predictions, while the least biased predictions are generated from stacked random forest ensemble.

In case of having crop-related APSIM variables as ML inputs, results indicate that stacked regression and stacked random forest ensembles make the best and the least biased predictions, respectively.

When the soil and weather-related APSIM variables are considered as ML inputs, the results show that stacked regression ensemble makes decent predictions with having the least amount of prediction error as well as bias.

*Table 4: Test set prediction errors of ML models for partial inclusion of APSIM variables (Test set is set to be the data for the year 2018)*

| *ML model* | *Phenology-related* | | | | *Crop-related* | | | | *Soil and weather-related* | | | |
|---|---|---|---|---|---|---|---|---|---|---|---|---|
| | *RMSE (kg/ha)* | *RRMSE (%)* | *MBE (kg/ha)* | *$R^2$ (%)* | *RMSE (kg/ha)* | *RRMSE (%)* | *MBE (kg/ha)* | *$R^2$ (%)* | *RMSE (kg/ha)* | *RRMSE (%)* | *MBE (kg/ha))* | *$R^2$ (%)* |
| LASSO | 1193 | 9.8% | -275 | 20.0% | 1148 | 9.4% | -466 | 26.1% | 1103 | 9.0% | -445 | 31.7% |
| XGBoost | 1221 | 10.0% | -655 | 16.3% | 1114 | 9.1% | -626 | 30.3% | 1036 | 8.5% | -445 | 39.8% |
| LightGBM | 1061 | 8.7% | -559 | 36.8% | 975 | 8.0% | -448 | 46.7% | 1052 | 8.6% | -515 | 37.9% |
| Random forest | 1562 | 12.8% | -1135 | -37.0% | 1208 | 9.9% | -796 | 18.0% | 1603 | 13.1% | -1195 | -44.2% |
| Linear regression | 1176 | 9.6% | 584 | 22.3% | 1014 | 8.3% | 193 | 42.3% | 965 | 7.9% | 445 | 47.7% |
| Optimized w. ens. | 1053 | 8.6% | -535 | 37.8% | 940 | 7.7% | -338 | 50.4% | 945 | 7.7% | -361 | 49.9% |
| Average ensemble | 1075 | 8.8% | -408 | 35.1% | 1002 | 8.2% | -429 | 43.6% | 998 | 8.2% | -431 | 44.1% |
| Stacked reg. ens. | 1040 | 8.5% | -574 | 39.3% | 906 | 7.4% | -234 | 53.9% | 837 | 6.8% | -121 | 60.7% |
| Stacked LASSO ens. | 1049 | 8.6% | -584 | 38.3% | 911 | 7.4% | -247 | 53.4% | 844 | 6.9% | -144 | 60.0% |
| Stacked Random f. ens. | 1228 | 10.0% | -134 | 15.4% | 1064 | 8.7% | -50 | 36.5% | 973 | 8.0% | -141 | 46.8% |
| Stacked LightGBM ens. | 1116 | 9.1% | -315 | 30.1% | 1032 | 8.4% | -83 | 40.2% | 958 | 7.8% | -150 | 48.5% |

Table 4 presents the test set prediction errors of designed ML models for all three scenarios of partial inclusion of APSIM variables. Overall results indicate that soil and weather-related APSIM variables as well as crop-related variables have a more significant influence on the predictions made by ML. This is interesting and is partially explained by the fact that ML somehow already accounts for phenology-related parameters, which are largely weather-driven, while the soil-related parameters are more complicated parameters that ML alone cannot see. This is more evident in Figure 6. Furthermore, it can be observed that some of the soil and weather-related ensemble models provide improvements over the models that we developed earlier including all APSIM variables. This result suggests that not all the included APSIM variables have useful information for ML yield prediction.

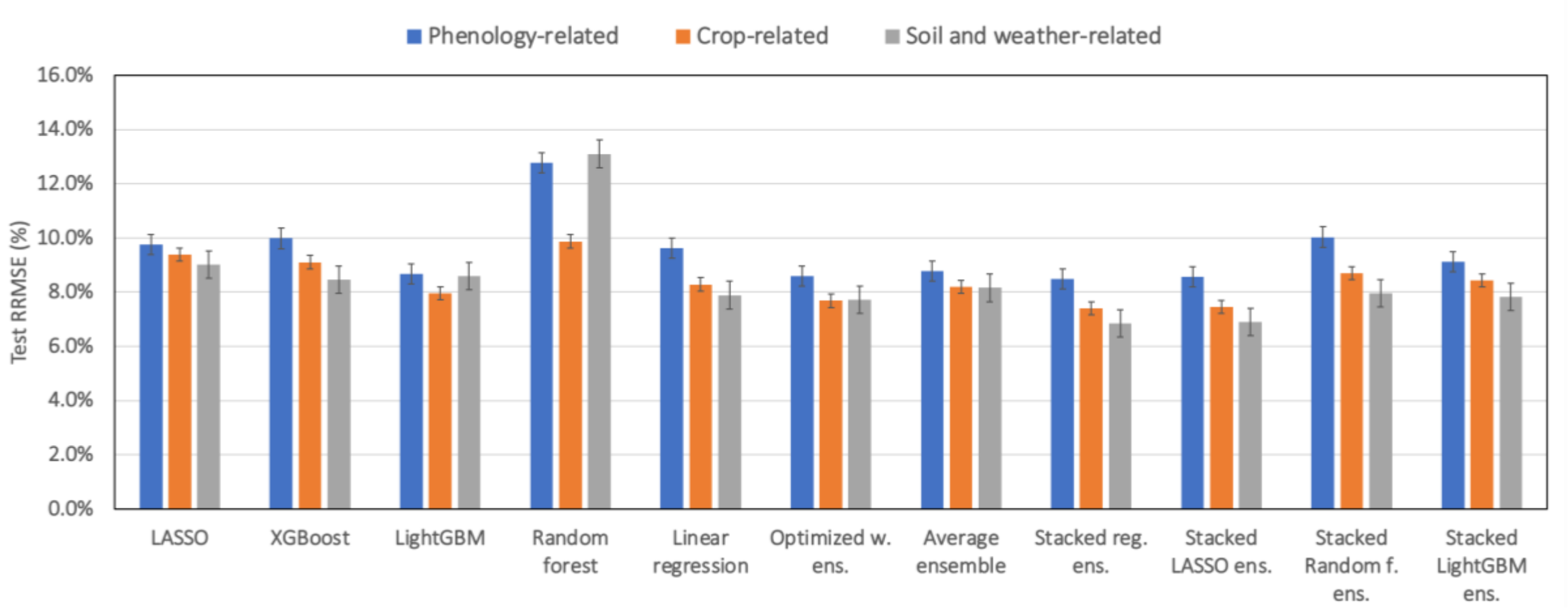

*Figure 6: Comparing test errors of three scenarios of partial APSIM variables inclusion. (Test data is set to be the data from the year 2018)*

### 3.4. Variable importance

The permutation importance (see also section 2.2.2) of five individual base models (linear regression, LASSO regression, XGBoost, LightGBM, and random forest) was calculated using the test data of the year 2018. Figure 8 depicts the top-15 normalized average permutation importance of these ML models. It should be noted that due to black-box nature of ensemble models, only individual learners were used to calculate permutation importance.

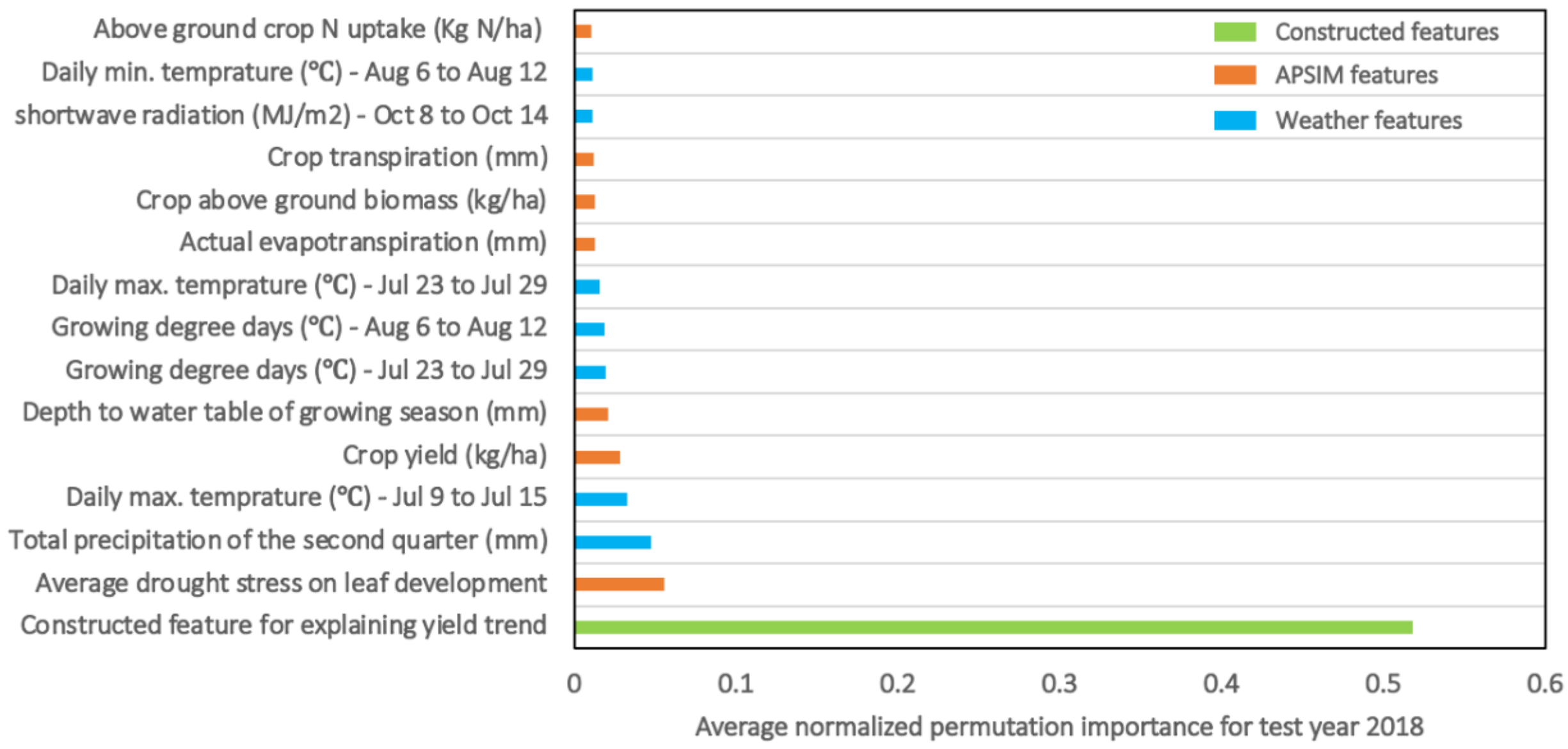

*Figure 7: Top-15 average normalized permutation importance of individual ML models for test year 2018*

Figure 7 indicates that the most important input feature for ML models is “yield_trend” which is the feature we constructed for explaining the increasing trend in corn yields and incorporated technological advances over the years (genetics and management improvement). Of the next 14 most important input features, seven variables were APSIM variables, while the remaining seven are weather input variables. Regarding the APSIM variables, five input features are part of crop-related APSIM variables, and the other two APSIM features are as soil and weather-related variables. This is in-line with the results of partial inclusion of APSIM variables discussed in section 3.3.

To find out which APSIM features have been more influential in predicting yields, the average permutation importance of five individual models (linear regression, LASSO regression, LightGBM, XGBoost, and random forest) was calculated for each test year. Figure 8 demonstrates the ranking of top 10 APSIM features. Results indicate that the AvgDroughtStress, AvgWTInseason, and CropYield were the most important features for machine learning models to predict yield. Most of these are water-related features suggesting the importance of soil hydrology in crop yield prediction in the US Corn belt. This result was consistent across three years, including the year drought 2012, in which model prediction was lower than the other years.

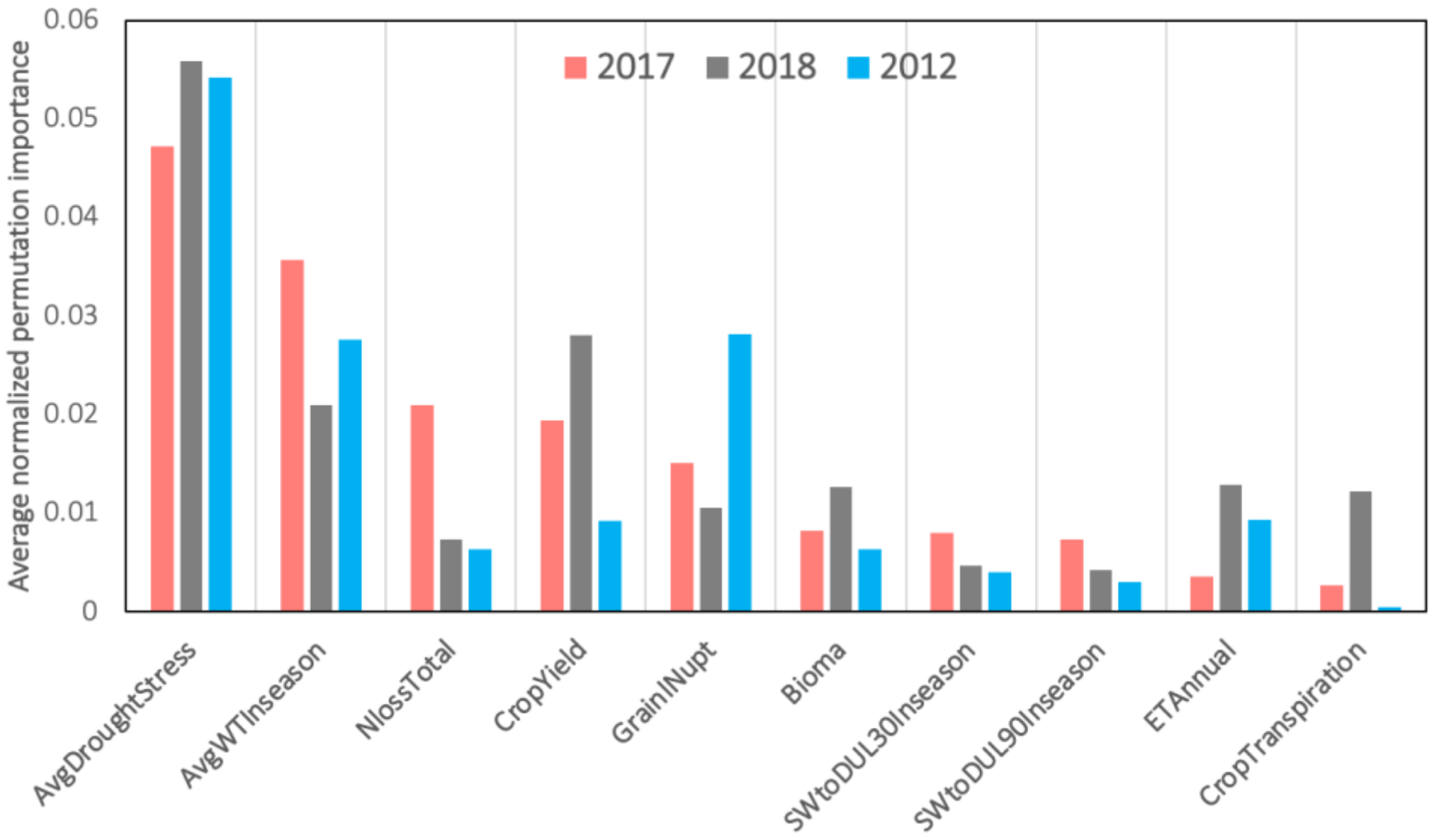

*Figure 8: Average normalized permutation importance of APSIM features for all test years. AvgDroughtStress: Average drought stress on leaf development, AvgWTInseason: Depth to water table of growing season (mm), NlossTotal: Total N loss (denitrification and leaching) (kg N/ha), CropYield: Crop yield (kg/ha), GrainINupt: Grain N uptake (kg N/ha), Bioma: Crop above ground biomass (kg/ha), SWtoDUL30Inseason: Growing season average soil water to field capacity ratio at 30 cm, ETAnnual: Actual evapotranspiration (mm), CropTraspiration: Crop transpiration (mm)*

## 4. Discussion

We proposed a hybrid simulation-machine learning approach that provided improved county-scale crop yield prediction. To the best of our knowledge, this is the first study that designs ensemble models to increase corn yields predictability. This study demonstrated that introducing APSIM variables into machine learning models and utilizing them as inputs to a prediction task on average can decrease the prediction error measure by RMSE between 7% and 20%. In addition, the predictions made by the hybrid model show less bias toward actual yields. Other studies in this area, are mainly limited in coupling simplest statistical models, i.e. linear regression variants, with simulation crop models and apart from two recent studies[35,36] there has been no study combining machine learning and simulation crop models. Considering the hybrid models, some of the developed models provided predictions with RRMSE values as small as 6-7%. This indicates that the developed models outperform the corn yield prediction models developed in the literature[24,72–75].

In addition to the prediction advantages achieved by coupling ML and simulation crop modelling, we investigated the value of different types of APSIM variables in the ML prediction and found out that soil water related APSIM variables contributed the most in improving yield prediction. The inclusion of APSIM consistently improved ML yield prediction in all years (2012, 2017, 2018). We also noticed that neither ML nor the hybrid model could sufficiently predict yields of the 2012 dry year. This suggests that more work is needed to adequately predict yields in extreme weather years, which are expected to increase with climate change[76–79], but we noticed that yield prediction of the dry year was better done by the hybrid model. Developing models that are more robust to extreme values, including additional climate information that can help the model to detect the drought, and including remote sensing data can be future research directions.

Designing a method that enables the ML models to capture the yearly increasing trends in corn yields was the main challenge of this work. To address this challenge, an innovative feature was constructed that could explains the trend to a great extent and as the variable importance results showed, it is by far the most important input feature for predicting corn yields.

The significant merits of coupling ML and simulation crop models shown in this study raise the question that whether the ML models can further benefit from addition of more input features from other sources. Hence, a possible extension of this study could be inclusion of remote sensing data into the ML prediction task and investigate the level of importance each data source can exhibit.

It should be also acknowledged that APSIM simulations that used as inputs to ML model leveraged the full weather of each test year. In real word applications, the weather will be unknown and the APSIM model would need to run in a forecasting mode[1,12,80] introducing some additional uncertainty. This is something to be explored further in the future.

## 5. Conclusion

We demonstrated improvements in yield prediction accuracy across all designed ML models when additional inputs from a simulation cropping systems model (APSIM) are included. Among several crop model (APSIM in this study) variables that can be used as inputs to ML, analysis suggested that the most important ones were those related to soil water, and in particular growing season average drought stress, and average depth to water table. We concluded that inclusion of additional soil water related variables (either from simulation model or remote sensing or other sources) could further improve ML yield prediction in the central US Corn Belt.

## Author Contribution Statement

MS is the lead author. He conducted the research and wrote the first draft of the manuscript. GH secures funding for this study, oversees the research, reviews and edits the manuscript. SA provides the data and guidance for the research. He also reviews and edits the manuscript. IH prepared the APSIM data.

## Legends

### Figures

Figure 1. Conceptual framework of this study's objective. This study investigates the effect of coupling process-based modeling with machine learning algorithms towards improved crop yield prediction.

Figure 2. Measured (USDA-NASS) corn yields vs. simulated corn yields at the state level from 1984 to 2019 using the pSIMS-APSIM framework.

Figure 3. Plotting aggregated annual yields for all locations under study and the average yields per year. The figure clearly shows a visible, increasing trend in yields. The blue line shows the yearly increasing trend in the yields.

Figure 4. Comparing average test RRMSE of benchmark and hybrid developed ML models. All developed models reveal superiority of hybrid models compared to the benchmark.

Figure 5. X-Y plots of some of the designed models for benchmark (top) and hybrid (bottom) cases for test year 2018. The intensity of the colors shows the accumulation of the data points.

Figure 6. Comparing test errors of three scenarios of partial APSIM variables inclusion (Test data is set to be the data from the year 2018).

Figure 7. Top-15 average normalized permutation importance of individual ML models for test year 2018

Figure 8. Average normalized permutation importance of APSIM features for all test years. Refer to Table 1 for explanation of the variables. AvgDroughtStress: Average drought stress on leaf development, AvgWTInseason: Depth to water table of growing season (mm), NlossTotal: Total N loss (denitrification and leaching) (kgN/ha, CropYield: Crop yield (kg/ha), GrainINupt: Grain N uptake (kg N/ha), Bioma: Crop above ground biomass (kg/ha), SWtoDUL30Inseason: Growing season average soil water to field capacity ratio at 30 cm, ETAnnual: Actual evapotranspiration (mm), CropTraspiration: Crop transpiration (mm)

### Tables

Table 1. Description of all APSIM outputs added to the developed data set for building ML models.

Table 2. Test set prediction errors of ML models for benchmark and hybrid cases.

Table 3. Test set prediction errors of ML models for benchmark and hybrid cases when considering an extreme weather year (2012)

Table 4. Test set prediction errors of ML models for partial inclusion of APSIM variables (Test set is set to be the data for the year 2018).